# Néel-type skyrmion in $WTe_2/Fe_3GeTe_2$ van der Waals heterostructure

Yingying Wu[1,†], Senfu Zhang[2,†], Gen Yin[1], Junwei Zhang[2], Wei Wang[3], Yang Lin Zhu[4], Jin Hu[5], Kin Wong[1], Chi Fang[6], Caihua Wang[6], Xiufeng Han[6], Qiming Shao[1], Takashi Taniguchi[7], Kenji Watanabe[7], Jiadong Zang[8], Zhiqiang Mao[4], Xixiang Zhang[2], Kang L. Wang[1,*]

[1]*Department of Electrical Engineering, University of California - Los Angeles, California 90095, USA.*

[2]*Physical Science and Engineering Division, King Abdullah University of Science and Technology, Thuwal 23955-6900, Saudi Arabia.*

[3]*Key Laboratory of Flexible Electronics & Institute of Advanced Materials, Jiangsu National Synergetic Innovation Center for Advanced Materials, Nanjing Tech University, Nanjing 211816, China.*

[4]*Department of Physics, Pennsylvania State University, University Park, PA 16802, USA.*

[5]*Department of Physics, University of Arkansas, Fayetteville, AR 72701, USA.*

[6]*Institute of Physics, Chinese Academy of Sciences, Beijing 100190, China.*

[7]*National Institute for Materials Science, 1-1 Namiki, Tsukuba 305-0044, Japan.*

[8]*Department of Physics and Astronomy, University of New Hampshire, Durham, New Hampshire 03824, USA.*

†These authors contributed to this work equally.

*Correspondence to: wang@ee.ucla.edu.

**The promise of high-density and low-energy-consumption devices motivates the search for layered structures that stabilize chiral spin textures such as topologically protected skyrmions. At the same time, layered structures provide a new platform for the discovery of new physics and effects. Recently discovered long-range intrinsic magnetic orders in the two-dimensional van der Waals materials offer new opportunities. Here we demonstrate the Dzyaloshinskii-Moriya interaction and Néel-type skyrmions are induced at the $WTe_2/Fe_3GeTe_2$ interface. $Fe_3GeTe_2$ is a ferromagnetic material with strong perpendicular magnetic anisotropy. We demonstrate that the strong spin-orbit interaction in 1T'-$WTe_2$ does induce a large interfacial Dzyaloshinskii-Moriya interaction at the interface with $Fe_3GeTe_2$ due to the inversion symmetry breaking to stabilize skyrmions. Transport measurements show the topological Hall effect in this heterostructure for temperatures below $100\,K$. Furthermore, Lorentz transmission electron microscopy is used to directly image Néel-type skyrmions along with aligned and stripe-like domain structures. This interfacial coupling induced Dzyaloshinskii-Moriya interaction is estimated to have a large energy of $1.0\ mJ/m^2$, which can stabilize the Néel-type skyrmions in this heterostructure. This work paves a path towards the skyrmionic devices based on van der Waals layered heterostructures.**

Atomically thin, layered van der Waals (vdW) materials have been experimentally shown to host long-range magnetic orders recently[1,2], which could push the magnetic memory and information storage to the atomically thin limit and lead to ultra-compact next-generation spintronics[3,4]. Since the discovery of the ferromagnetism in $Cr_2Ge_2Te_6$[5,6], $CrI_3$[7-10] and $Fe_3GeTe_2$ (FGT)[11-14], such layered crystals have been at the frontier of material research. Besides the material itself, interfacial engineering in vdW heterostructures offers an effective methodology to spin-polarize or valley-polarize 2D materials. Proximity effect from the interface has been widely researched for spin or valley polarization in 2D materials, like graphene on transition metal dichalcogenides (TMDs)[15-19] and $WSe_2$ on $CrI_3$[20-23]. Coupling 2D magnets to vdW materials not only lends the magnetic properties of these 2D magnets to the adjacent materials, but also modifies the magnetic properties of the 2D magnets themselves[24]. Among all the possible interfacial coupling for 2D magnets, spin orbit coupling proximity can play an important role when the atoms of 2D magnets are in contact of heavy elements, considering that the magnetic properties are intrinsically related to the spin orbit coupling.

Rashba spin-orbit coupling is known to lead to a strong Dzyaloshinskii-Moriya interaction (DMI) at the interface[24,25], where the broken inversion symmetry at the interface can change the magnetic states. DMI has been recognized as a key ingredient in the creation, stabilization and manipulation of skyrmions and chiral domain walls. Whereas skyrmions from DMI become significant in the heavy metals/ferromagnet systems[26–28], there have been no **direct** observations of skyrmions in van der Waals heterostructures, even though the topological Hall effect has been reported in Cr-doped topological insulator (TI)/TI[29] and Mn-doped TI systems[30].

In this work, we have observed the topological Hall effect in $WTe_2$/FGT vdW heterostructures from transport measurements. More importantly, a large DMI energy of $1.0\ mJ/m^2$ has been determined in this system and the formation of Néel-type skyrmions has been captured with Lorentz transmission electron microscopy (L-TEM). The sizes of the directly observed skyrmions are ~$150\ nm$ at $94\ K$ and ~$80\ nm$ at $198\ K$. This work helps promote 2D materials for ultra-compact spintronic devices.

**Thickness characterization of $WTe_2$ and FGT**

Before we show the proximity effect of the heterostructures, we first determine the properties of $WTe_2$ and FGT separately. The 1T' crystal structure of $WTe_2$ (Fig. 1a) has been confirmed in the previous work[31]. Atomic force microscopy was adopted to measure the thickness of this material as shown in Figs. 1 b-c. It shows that the 1L $WTe_2$ on the $SiO_2$/Si substrate has a height of about 1.1 $nm$. The transport data of 1L $WTe_2$ are given in the Supplementary Information Sec. A. Bulk FGT consists of weakly bonded $Fe_3Ge$ layers that alternate with two Te layers with a space group $P6_3mmc$ as shown in Fig. 1a (bottom). For FGT, the 1L thickness is $0.8\,nm$[11]. This thickness is used to determine the number of layers for FGT as shown in Figs. 1d-e. As shown in Figs. 1f-i, the Curie temperature decreases from ~$200\,K$ to ~$100\,K$ when the thickness of FGT goes down from 60L to 4L. The perpendicular magnetic anisotropy is well preserved for FGT with thickness down to seven layers. For the thin FGT samples with seven layers and four layers, hexagonal boron nitride (h-BN) thin flakes are used for protection from the ambient.

## Topological Hall effect in $WTe_2$/FGT heterostructure

$WTe_2$ has one of the largest spin-orbit coupling among the transition metal dichalcogenides[32]. At the interface, the terminated atoms in $WTe_2$ are the same as those in FGT: two layers of heavy Te atoms are coupled through 5p orbital coupling, where the strong spin-orbit interaction from $WTe_2$ could play a significant role in reorganizing the spin polarizations in FGT. Such effect can be captured by transport measurements in thin films, and the transport data of h-BN/$WTe_2$/FGT heterostructures are shown in Fig. 2 with h-BN serves as the protection layer (device fabrication details could be found in Supplementary Information Sec. B). For the h-BN/1L $WTe_2$/4L FGT sample (sample A) as shown in Figs. 2a-b, the longitudinal resistivity increases when temperature goes down from room temperature. The Curie temperature for this device is below $150\,K$, which is close to that of a 4L FGT on $SiO_2$/Si sample. Topological Hall is a spin-chirality-driven Hall effect, i.e., the spin chirality induces a finite contribution to the Hall response. In our case, dips and peaks near the magnetic phase transition edge, which signal the presence of the topological Hall effect, show up below $100\,K$ as shown in Fig. 2c. For example, at $50\,K$ there is a dip or peak at a field strength $\sim \pm\ 975\ Oe$ as indicated by the dashed lines. Different from this sample where the interfacial coupling has been reflected through the topological Hall effect, a 2L $WTe_2$/30L FGT heterostructure (sample B) shows the perpendicular magnetic anisotropy is well preserved in FGT when the temperature is below $180\ K$ in the Fig. 2e, indicated by the square loop of the Hall resistivity. However, within an intermediate temperature range ($180\ K$ -$200\ K$), the hysteresis loops differ in having transitions in steps. For example, at $190\ K$, the Hall resistivity suddenly jumps from the low saturation value to an intermediate level and then changes linearly. Finally, it saturates to the high saturation value at a relatively high positive field. The reversed trace does the opposite as expected. A possible explanation for this behavior is the formation of labyrinthine domain structures in FGT which results in multi domains[12], and this will be confirmed later. Compared to the sample A as shown in Figs. 2a-c, the sample B as shown in Figs. 2d-e with a much thicker FGT flakes measures the averaged transport signal which may lead to the smearing out of the topological Hall signal at the interface and will be discussed later. In the following part, we utilized L-TEM to directly investigate the domain structure in 2L $WTe_2$/30L FGT heterostructure, which is similar to sample B and confirm a large DMI at the interface.

**Experimental observation of Néel-type skyrmions with L-TEM**

Lorentz microscopy methods allow us to obtain direct magnetic domain structural images of thin magnetic films based on the fact that electrons experience a Lorentz force when traveling in a magnetic field. Compared to other techniques[33], L-TEM, as one of the most direct methods to observe the magnetic domain structures, domain walls and skyrmions, affords the advantage of a spatial resolution below $5\ nm$. The contrast formed in L-TEM is traditionally explained from the deflection of electrons due to the Lorentz force. For materials like FGT with a perpendicular magnetic anisotropy, the magnetization of the sample (at saturation) is parallel to the electron velocity direction. This suppresses the electron deflection, resulting in zero contrast. Thus sample plane was tilted to have partially the in-plane magnetization for detecting the Néel-type skyrmion as schematically shown in Fig. 3a. The resulted image for a skyrmion should be of dark-bright contrast due to the contributions both from the outside and core of the skyrmion[34]. The skyrmion lattice was observed in another 2L $WTe_2$/40L FGT heterostructure at $180\ K$ for the rotating angle $\alpha$ = 30° (Fig. 3b) in de-focused images. From under focus to over focus, the skyrmions transform from white on the top and dark in the bottom to the opposite contrasts. These are consistent with the Néel-type nature of the skyrmions. One hexagonal skyrmion lattice is indicated with the dashed lines. The field dependence of a single skyrmion in the $WTe_2$/30L FGT (similar to sample B for the transport) is shown in Fig. 3c when a field changes from $540\ Oe$ to $660\ Oe$ at $94\ K$. The Néel-type skyrmion is well developed at a field of $540\ Oe$ and $600\ Oe$ along the z direction, having dark on the top side and bright in the bottom. The size is estimated to be $\sim 150\ nm$ (more information about the skyrmion size of $\sim 150\ nm$ at $100\ K$ and $\sim 80\ nm$ at $197\ K$ can be found in Supplementary Information Sec. H). However, we failed to resolve any domain structure in 1L $WTe_2$/4L FGT samples (similar to sample A) using L-TEM. This may be due to the small sheet magnetization in the thin film, which requires higher beam exposure to resolve; however, this already exceeds the tolerance of our samples, beyond which, unrecoverable damages occur. L-TEM measurements of $WTe_2$/FGT heterostructures with varied FGT thicknesses suggests the DMI only penetrates to a finite depth of FGT (more details shown in Supplementary Information Sec. F). To further understand how this interfacial DMI penetrates through the FGT layers, we have carried out simulations as shown in Fig. 3d. For FGT away from the interface, it enters the ferromagnetic phase as shown in the vertical profile in the $yz$ plane.

**Experimental evidence of $WTe_2$ induced DMI**

The strong perpendicular magnetic anisotropy favors out-of-plane magnetization, which makes the spin polarization in the FGT form the domain in the up or down directions. Fig. 4b shows the magnetic domain images for a 30L FGT thin flake without $WTe_2$. The film exhibits labyrinth domains at $94\ K$ and $0\ T$. When the in-plane field is tuned to the opposite direction, the contrast of domain edge completely switches the sign as shown in the cases of a tilt angle of $\alpha$= -20° and $\alpha$=21° as shown in Fig. 4b.

The formation and structure of domain walls are usually a result of the interplay between exchange interaction, magnetic anisotropy and dipolar interaction. Reducing the thickness of the FGT flakes decreases the dipolar interaction, thus, perpendicular magnetic anisotropy leads to a stabilized single domain[14]. Meanwhile, the exchange interaction term here includes DMI, which is an antisymmetric exchange interaction that favors a chirally rotating magnetic structure of a specific rotational direction. By comparing the $WTe_2$/FGT and nearby FGT regions, the magnetic domain difference shows the interfacial coupling contributes to a strong DMI.

In the $WTe_2$/FGT heterostructure, there is no contrast when $\alpha = 0°$ as shown in Fig. 4c, which is an evidence for the Néel-type domains. When $\alpha \neq 0°$, we have observed this well aligned and stripe-like domains with a domain width $w = 290\ nm$ at $0\ T$, which is sharply different from the domain structure of the FGT flake as shown in Fig. 4b, but with a much smaller domain width. Thus the DMI is largely enhanced[28] for this heterostructure from the interfacial Te atoms coupling. A phenomenological model defines the dependence of the domain width $w$ on the domain wall energy $\delta_W$ by[35]:

$$w = \beta \frac{4\pi\delta_W}{M_S^2}, \tag{1}$$

here the domain wall energy $\delta_W$ is related to exchange stiffness $A$, effective anisotropy constant $K_{eff}$ and DMI constant $|D|$ by $\delta_W = 4\sqrt{AK_{eff}} - \pi|D|$ [36]. $\beta$ is a phenomenological fitting parameter and taken to be 0.31 for FGT[35]. A domain width of $290\ nm$ leads to a domain wall energy $\delta_W = 0.77\ mJ/m^2$ including the DMI term. In the case of pristine FGT without considering DMI, the domain wall energy is simply expressed as $4\sqrt{AK_{eff}}$. Then by comparing these two domain wall energies with and without the DMI contribution, we obtain a DMI energy $|D| = 1.0\ mJ/m^2$ in our system, which is comparable to the previous value in heavy metal/ferromagnet thin film systems[27] (The detailed estimation of this $|D|$ can be found in the Supplementary

Information Sec. J). One can compare $|D|$ to the critical value $|D_c|$, which is required to stabilize chiral Néel domain wall[27,37],

$$|D_c| = \frac{4}{\pi}\sqrt{\frac{A}{K_{eff}}}K_d, \tag{2}$$

where when the magnetostatic or stray field energy constant $K_d = 2\pi M_s^2$ is large, critical value for the Néel type domain is larger. Thus, $|D_c|$ is calculated to be $\sim 0.1\ mJ/m^2$, so that $|D| > |D_c|$ and chiral Néel textures are expected. Compared to the transport measurements which takes the averaged transport signal of the interfacial few nanometers FGT coupled to $WTe_2$ near the interface and other FGT layers away from the interface, L-TEM helps confirm the DMI and Néel-type skyrmion at the interface for the $WTe_2$ with FGT samples (check Supplementary Information Sec. E for more details).

However, to bridge the gap of different thicknesses between the transport and the L-TEM studies, here we show two results suggesting good consistency. On one hand, both skyrmions from the topological Hall effect and the L-TEM images give the sizes within an order of magnitude (details could be found in Supplementary Information Sec. K). On the other hand, the $WTe_2$ capping can only impact the domain structure for the FGT thickness <65L, suggesting a finite vertical penetration depth of the DMI. Assuming an exponential decay in the DMI profile, our simulation suggests that the skyrmions can only penetrate to a finite depth, where a large volume of the ferromagnetic phase shows up away from the interface. Due to frequent scatterings, when carriers are passing through the ferromagnetic phase, they may quickly lose the memory of the transverse velocity due to the topological Hall effect, and therefore the anomalous Hall effect dominates. This may explain the missing topological Hall resistivity humps in thicker films.

In summary, we reported the observed Néel-type skyrmions from L-TEM and the discovery of topological Hall effect in the $WTe_2$/FGT heterostructure. A large DMI energy of $\sim 1.0\ mJ/m^2$ from the interfacial coupling may be a result of the broken inversion symmetry from Rashba spin-orbit coupling. We have shown Néel-type skyrmions of a small size ($\sim 150\ nm$ at $94\ K$ and $\sim 80\ nm$ at $198\ K$) at the interface of the vdW heterostructure. Further researches can include electrically gating to control skyrmions in 2D vdW heterostructures, and this may open a new area in the field of ultra-compact next-generation spintronics.

**Methods**

**Sample assembly using a pick-up transfer technique:** In preparation, we coated the polydimethylsiloxane (PDMS) stamp on a glass slide with a polypropylene carbonate (PPC). During the assembly of heterostructure, we first exfoliated h-BN onto the $300$-$nm$-thick $SiO_2$/Si substrate. Then the PDMS/PPC on the glass slide was used to pick up the h-BN from the $SiO_2$/Si substrate by heating up to around $40°C$. After that, we aligned the PDMS/PPC/h-BN and the exfoliated $WTe_2$ on the $300$-$nm$-thick $SiO_2$/Si substrate, and used the similar heating temperature to pick up $WTe_2$. Then we used PDMS/PPC/h-BN/$WTe_2$ to pick up FGT. By heating the samples up to $110°C$, PPC was released from the PDMS and the PPC/h-BN/$WTe_2$/FGT was transferred onto the prepared bottom electrodes. Then acetone was used to remove the PPC thin film. The metal contacts used for the transport measurements of FGT thin flakes have two types, one is evaporation of Cr/Au electrodes after exfoliation of FGT onto the $SiO_2$/Si substrates and the other one is using the pick-up transfer technique to accurately align FGT with the prepared bottom Cr/Au electrodes. The length and width of contacts in a Hall bar geometry as shown in Fig.1f-i for 60L FGT, 30L FGT, h-BN/7L FGT, h-BN/4L FGT are $15\ \mu m \times 9\ \mu m$, $15\ \mu m \times 9\ \mu m$, $3.5\ \mu m \times 0.5\ \mu m$, and $3.5\ \mu m \times 0.5\ \mu m$,, respectively.

**L-TEM measurements**: The in-situ L-TEM imaging was carried out by using a FEI Titan Cs Image TEM in Lorentz mode (the Fresnel imaging mode) at $300\ kV$.

**Acknowledgements**

The sample synthesis and characterization efforts are supported by the US Department of Energy under grand DE-SC0019068. The transport measurements in this work are supported by the ARO program under contract W911NF-15-1-10561, the Spins and Heat in Nanoscale Electronic Systems (SHINES), an Energy Frontier Research Center funded by the US Department of Energy (DOE), Office of Science, Basic Energy Sciences (BES) under award #SC0012670. We are also grateful to the support from the National Science Foundation (DMR-1411085). L-TEM measurements is based on research supported by the King Abdullah University of Science and Technology, Office of Sponsored Research and under the award No. OSR-2016-CRG5-2977.

**Contributions**

Y. W. and K. L. W. conceived the work and K. L. W. supervised the project. W. W. grew the FGT bulk crystals and Y. L. Z. and J. H. grew the $WTe_2$ bulk crystals. G. Y. did the simulation. T. T and K. W provided the h-BN crystal. Y. W. fabricated devices and performed transport measurements. S. Z. and J. Z. performed the Lorentz transmission electron microscopy measurements. K. W. and Y. W. carried out the atomic force microscopy measurements. C. F., C. W. and X. H. prepared the bottom electrodes. Y. W. and K. L. W. wrote the manuscript with input from all authors.

**Data Availability**

The data that support the findings of this study are available from the corresponding author upon reasonable request.

**Competing interests**

The authors declare no competing interests.

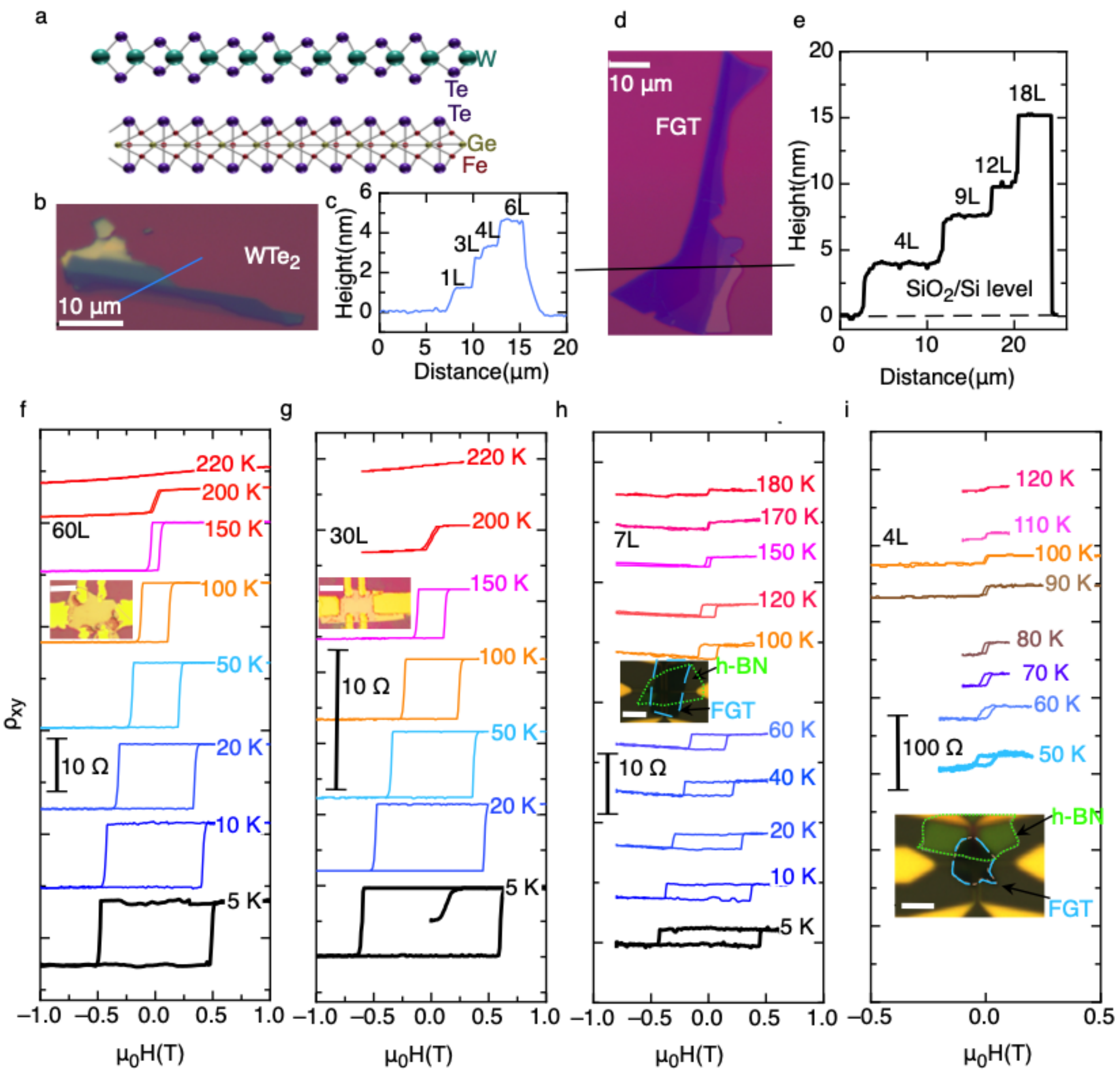

**Figure 1: Thickness characterization of $WTe_2$ and FGT flakes using atomic force microscopy (AFM) and layer-dependent transport properties of FGT flakes. a**, Schematic graph for $WTe_2$ on FGT. **b**, Microscopic image of exfoliated $WTe_2$ flakes. **c**, Cross-sectional profile of the $WTe_2$ flakes along the blue line shown in **b**. **d**, Microscopic image of exfoliated FGT thin films. **e**, Cross-sectional profile of the FGT flakes along the black line shown in **d**. Temperature dependence of Hall resistivity for **f**, 60L, **g**, 30L, **h**, 7L, and **i**, 4L FGT flakes showing that the Curie temperature decreases as the thickness of FGT decreases. Insets show the devices for the measurements separately. Scale bar in the inset: 10 $\mu m$. Resistivity is shifted for clarity.

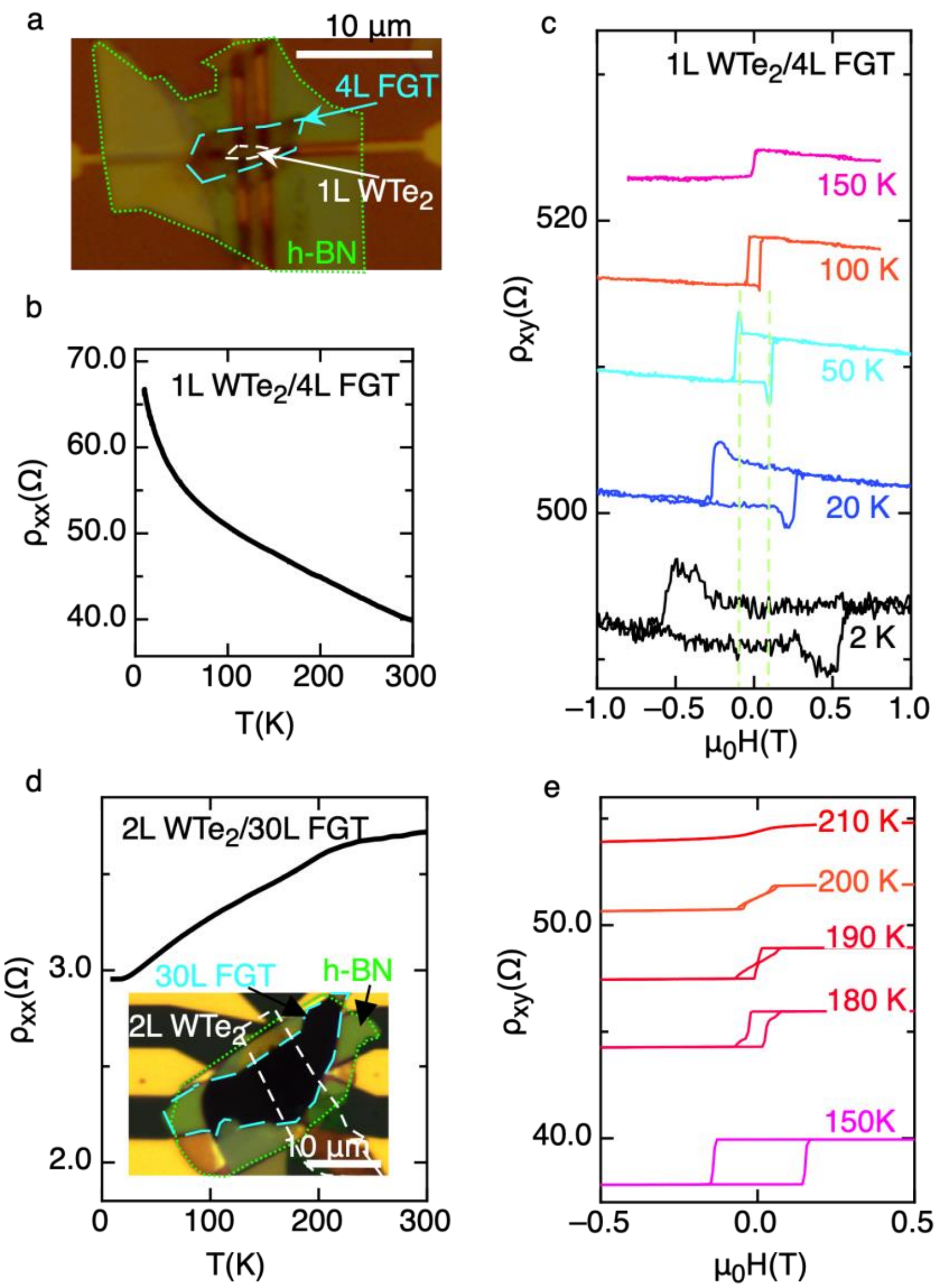

**Figure 2: Transport measurements of $WTe_2$/FGT heterostructures**. **a**, Microscopic image of sample A. **b**, Longitudinal resistivity dependence on the temperature shows the $\rho_{xx}$ increases when temperature drops. **c**, Hall resistivity of the heterostructure shown in **a**. Transverse resistivity shows a peak and dip near the transition edge before the magnetization saturates, which is a sign of the topological Hall effect. An offset is used for clarity. **d**, Longitudinal resistivity dependence on the temperature shows the metallic behavior when temperature drops. Inset shows the microscopic image of sample B. **e**, Hall resistivity of the heterostructure shown in **d**. An offset is used for clarity.

$H$ along out-of-plane direction.

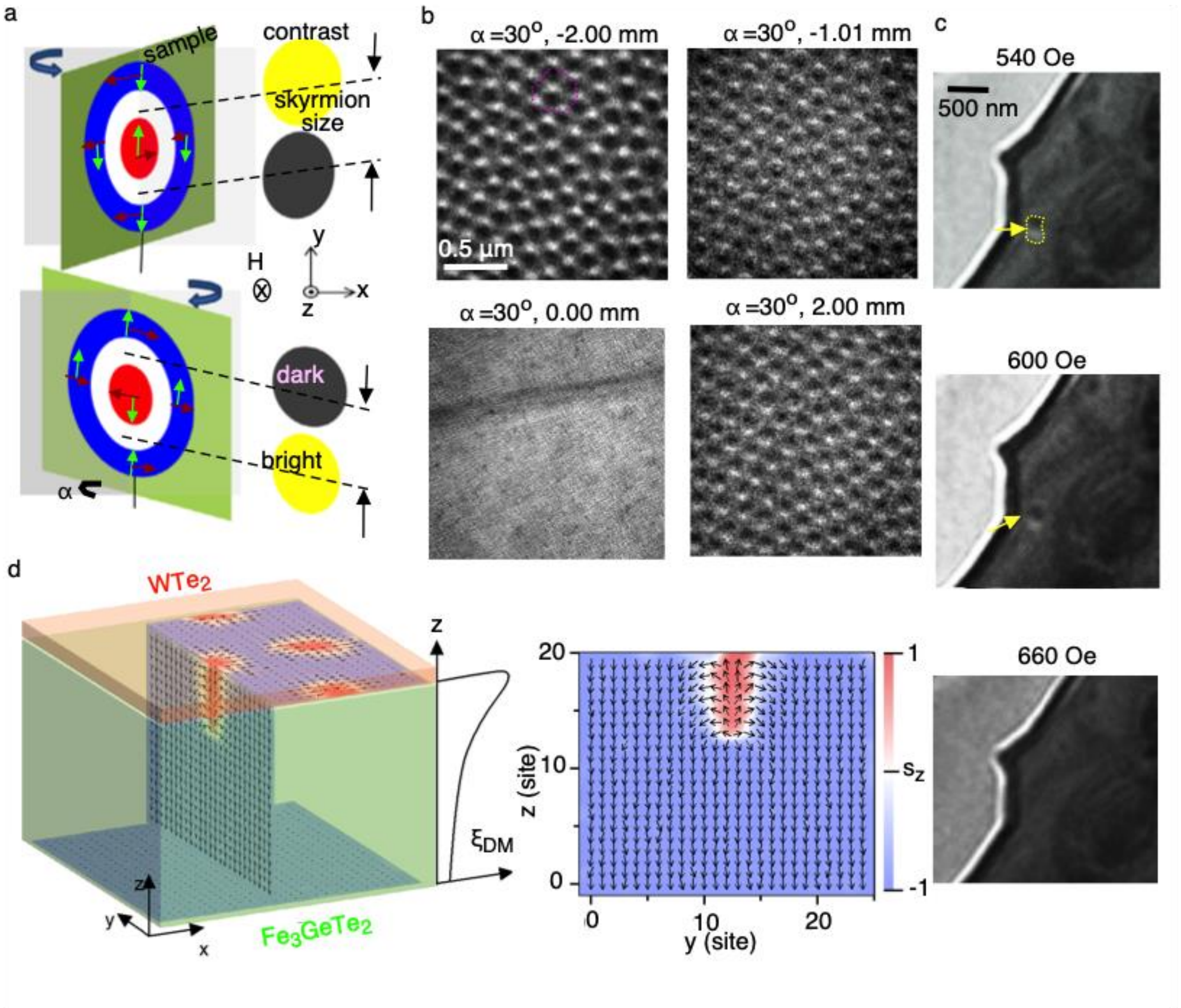

**Figure 3: Néel-type skyrmions observed under L-TEM. a**, Schematic diagram of a Néel-type skyrmion on a tilt sample for L-TEM imaging. The red and blue circles are for positive and negative magnetizations along $z$, respectively. Brown arrows indicate the in-plane magnetization component while green arrows indicate the Lorentz force. **b**, L-TEM observation of skyrmions lattice changed from under focus to over focus on $WTe_2$/40L FGT samples at 180 $K$ with a field of 510 $Oe$. **c**, L-TEM observation a Néel-type skyrmion at $T = 94\ K$, $\alpha = 21.86°$ and $H = 540\ Oe, 600\ Oe$, where $\alpha$ is the angle between the sample plane and $x$-$y$ plane. The yellow arrow points to a skyrmion. The skyrmion size is $\sim 150\ nm$. **d**, Simulation results considering a finite depth of interfacial DMI in FGT.

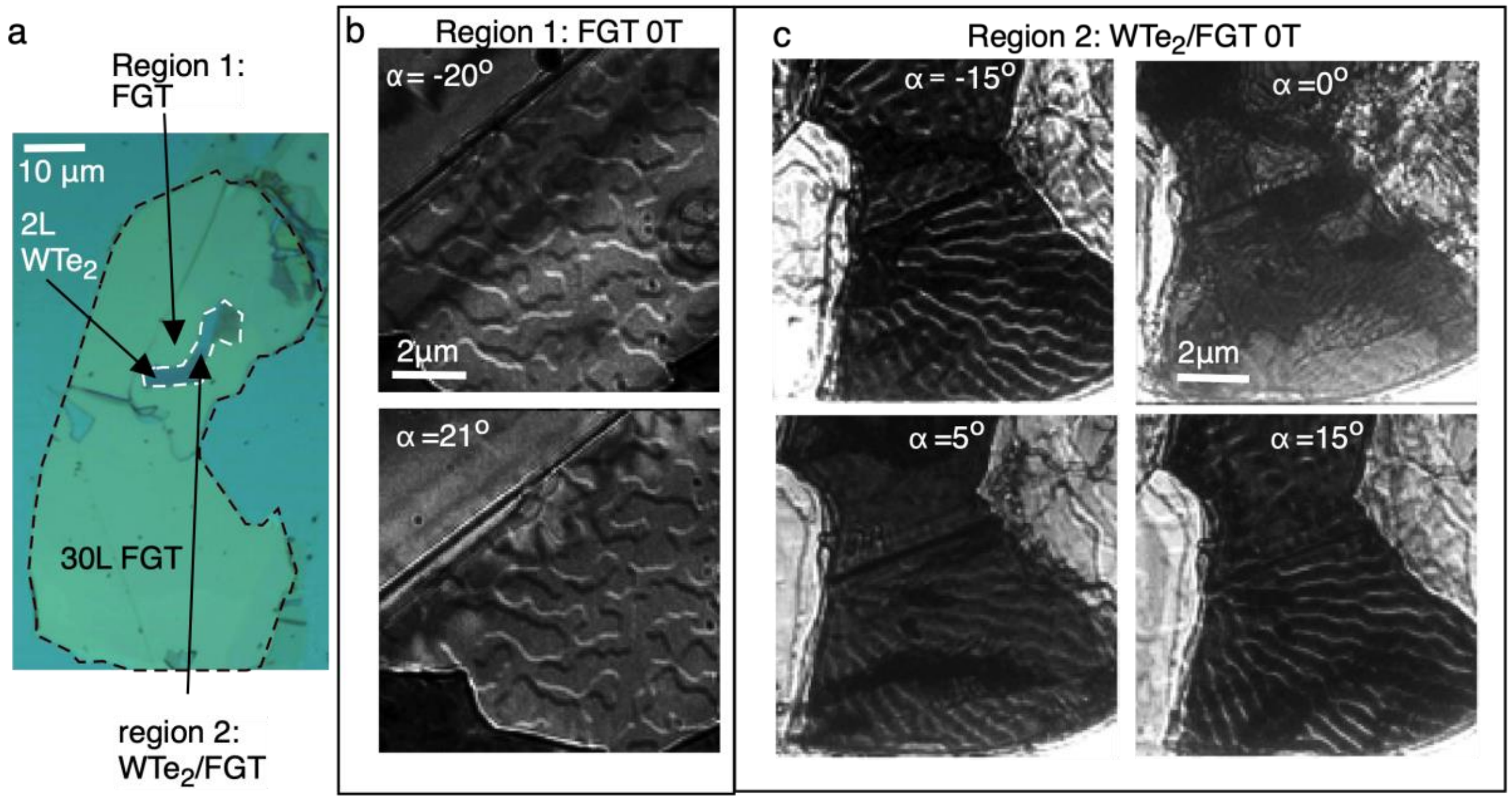

**Figure 4: Magnetic domains of FGT (region 1) and $WTe_2$/FGT heterostructures (region 2) with different tilt angles $\alpha$. a**, Sample for L-TEM consisting of 2L $WTe_2$ and 30L FGT. **b**, Typical labyrinth domain in 30L FGT thin flakes. **c**, From the aligned and stripe-like domain structures of the $WTe_2$/FGT, a DMI energy is estimated to be $\sim 1.0\ mJ/m^2$.

Supplementary Information for

# Néel-type skyrmion in $WTe_2$/$Fe_3GeTe_2$ van der Waals heterostructure

Yingying Wu[1, †], Senfu Zhang[2,†], Gen Yin[1], Junwei Zhang[2], Wei Wang[3], Yang Lin Zhu[4], Jin Hu[5], Kin Wong[1], Chi Fang[6], Caihua Wang[6], Xiufeng Han[6], Qiming Shao[1], Takashi Taniguchi[7], Kenji Watanabe[7], Jiadong Zang[8], Zhiqiang Mao[4], Xixiang Zhang[2], Kang L. Wang[1,*]

## A. Hall resistivity of monolayer $WTe_2$

To confirm the dip and peak near the magnetic transition edge in the transport signal as shown in Fig. 2 in the main text are from the topological Hall effect, we have shown the Hall resistivity of h-BN/monolayer $WTe_2$ in Fig. S1. This helps exclude the possibility that the dip and peak are from the transport signal of $WTe_2$ and thus leads to the conclusion that topological Hall effect exists at the interface.

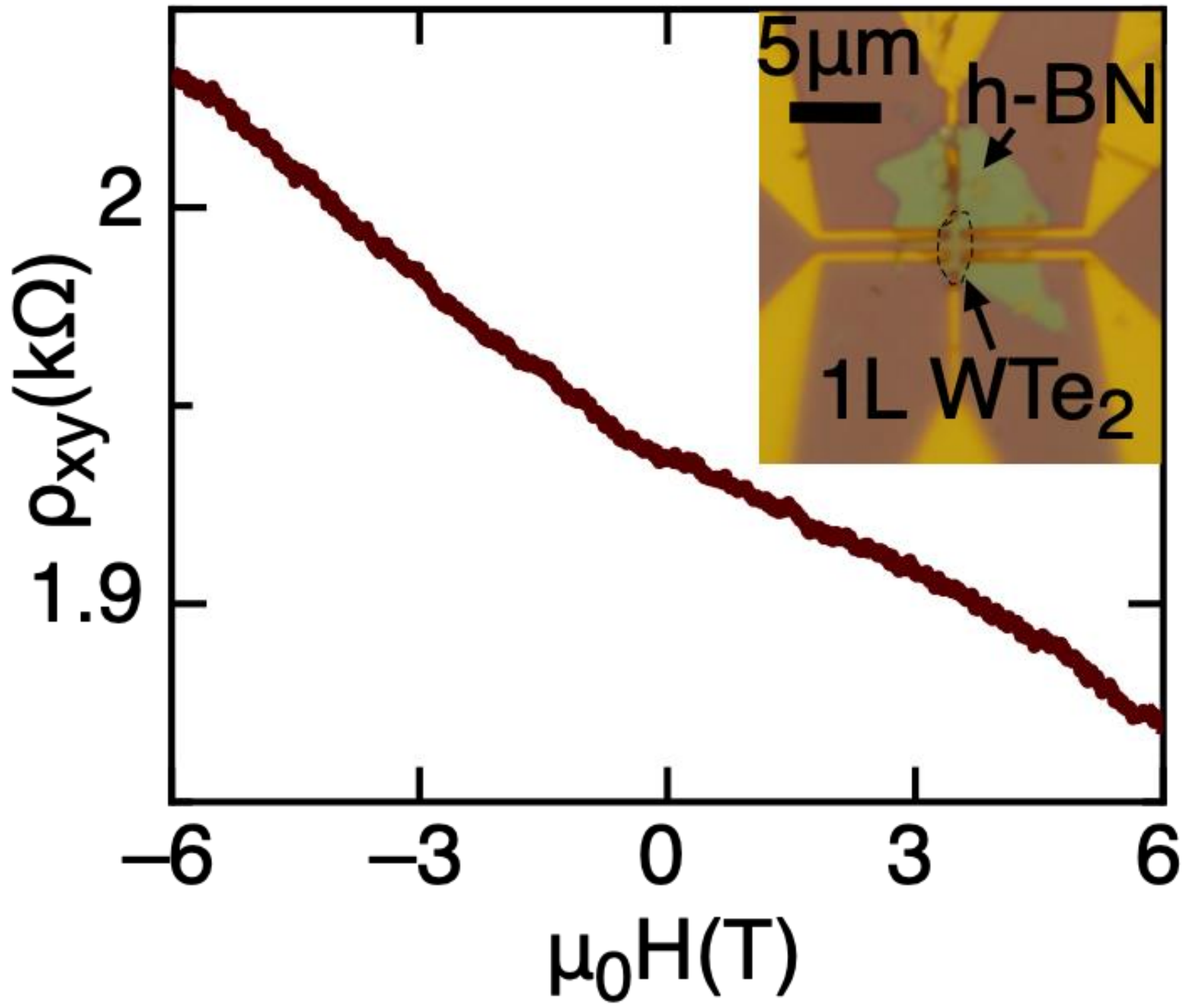

Fig. S1: Hall resistivity of a monolayer $WTe_2$ covered by h-BN on bottom electrodes at $2\ K$.

## B. Device fabrication

We've prepared the bottom electrodes by e-beam lithography first. Then $5\ nm$ Cr/$30\ nm$ Au was evaporated to form bottom electrodes. Then we exfoliated $WTe_2$ and FGT from high-quality bulk materials separately onto the $300\ nm$ $SiO_2$/Si substrates. After that, PDMS/PPC on a glass slide was used to pick up the monolayer or bilayer $WTe_2$ on the substrate. The pick-up procedure was to heat the sample stage up to $50°C$ when PDMS/PPC was lowered to touch $WTe_2$ and shut down the heating while detaching the PDMS/PPC from the sample stage. After the separation, the $WTe_2$ would be picked up by PDMS/PPC. Then PDMS/PPC/$WTe_2$ was used to pick up FGT thin layers. The resulted PDMS/PPC/$WTe_2$/FGT was then transferred onto the prepared bottom metal electrodes with proper alignment. After removing the PDMS/PPC by the acetone, $WTe_2$/FGT heterostructures are in good contact with the bottom electrodes. Finally, we always transfer h-BN thin layers onto this structure for protection. All the procedures were carried out inside a glove box, with $H_2O$ of 1.2 ppm (parts per million) and $O_2$ less than 50 ppm.

**C. ρ-$T$ curves for FGT with varied thickness**

As the thickness becomes smaller, the resistivity dependence on the temperature of FGT films changes from metallic to insulating when exfoliated on top of sapphire substrate, as has been shown before[1]. To confirm this transition, we have fabricated additional samples and carried out similar measurements in our $WTe_2$/FGT heterostructures. Here, Fig. S2a demonstrates the $\rho$-$T$ behavior for a 30L FGT capped with a 2L $WTe_2$. Such metallic $\rho$-$T$ behavior is mainly contributed by the 30L FGT layer, as shown by the control sample without the $WTe_2$ capping (Fig. S2b).

As the thickness of FGT goes down to 3L, the heterostructure becomes much insulating, as shown in Fig. S2c. As a comparison, we obtained the $\rho$-$T$ measurement for a monolayer $WTe_2$, which is roughly two orders of magnitude more insulating than the heterostructure. We therefore conclude that the carrier transport in Fig. S2c is also dominated by the FGT layer, which is not only $\sim$2 orders of magnitude more insulating than the case of a 30L FGT, but also presents a semiconducting $\rho$-$T$ trend. These observations are consistent with existing results, where uncapped FGT films were measured on $Al_2O_3$, as shown in Fig. S2e.

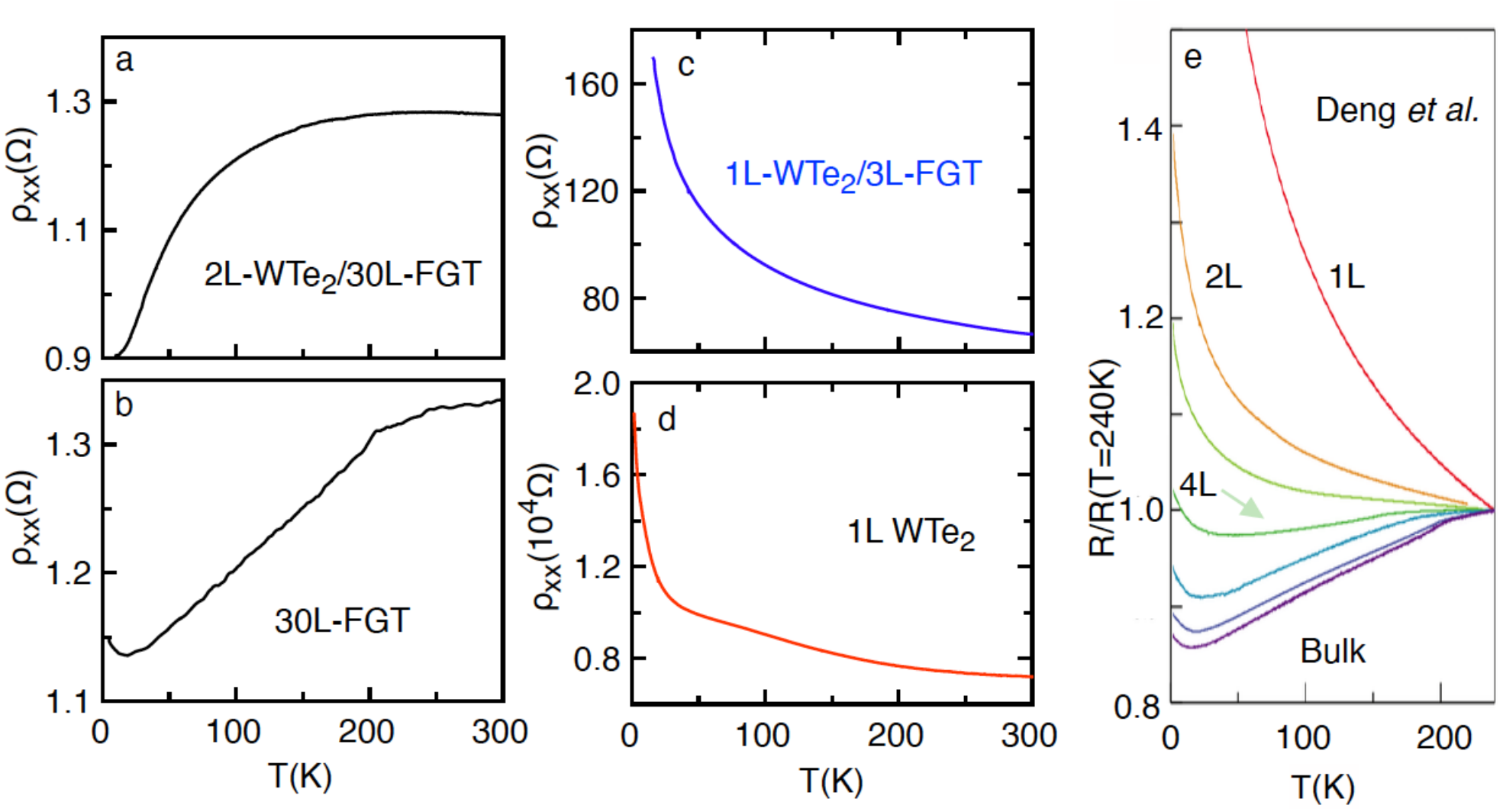

Fig. S2: Temperature dependence of $\rho_{xx}$ for $WTe_2$/FGT heterostructures. a, The $\rho_{xx}$-$T$ curve for a 30L FGT film with a 2L $WTe_2$ capping. b, The $\rho_{xx}$-$T$ result of the control sample, where a bare 30L FGT film is not capped with $WTe_2$. c, The $\rho_{xx}$-$T$ behavior of a 3L FGT thin film with a 1L $WTe_2$ capping. d, The $\rho_{xx}$-$T$ for a control sample with a 1L $WTe_2$ film. e, Previous $R_{xx}$-$T$ measurements of FGT films on $Al_2O_3$ thin films done in Ref. [1].

## D. Antisymmetric $\rho_{xx}$-$B$

For the 1L $WTe_2$/4L FGT heterostructure, the topological Hall signal from transverse resistivity has been shown in Fig. 2c in the main text. However, we also observed an antisymmetric magnetoresistivity (MR) as shown in Fig. S3a. Indeed, the antisymmetric MR behavior during the magnetic reversal could originate from the electron scattering due to magnetic domain walls in a thin-film magnet with perpendicular anisotropy[2]. If this were true, any misalignment of the transverse electrodes would capture such antisymmetric signal in $\rho_{xy}$ even without skyrmions. To rule out such alternative interpretation, we performed additional measurements, as discussed in the following.

If the antisymmetric behavior in $\rho_{xx}$ came from a domain wall, switching measuring electrodes from $\rho_{23}$ to $\rho_{65}$, the antisymmetric behavior would have changed polarity[2]. However, this was not the case. The antisymmetric behavior maintained its polarity, namely, a dip on the left and a spike on the right in both cases, as shown in Fig. S3b. This rules out the domain wall interpretation.

Indeed, the asymmetric $\rho_{xx}$ captured in our measurement is likely due to the mixing between $\rho_{xy}$ and $\rho_{xx}$. Such mixing occurs in many 2D material studies since it is difficult to control the geometry of exfoliated van der Waal materials. As shown in Fig. S4, the magnetoresistivity of a 60L FGT (on a $SiO_2$/Si substrate) also possesses hysteresis of ~0.1 Ω. Since it is technically difficult to separate longitudinal and transverse components, we show the unsymmetrized raw data throughout the manuscript.

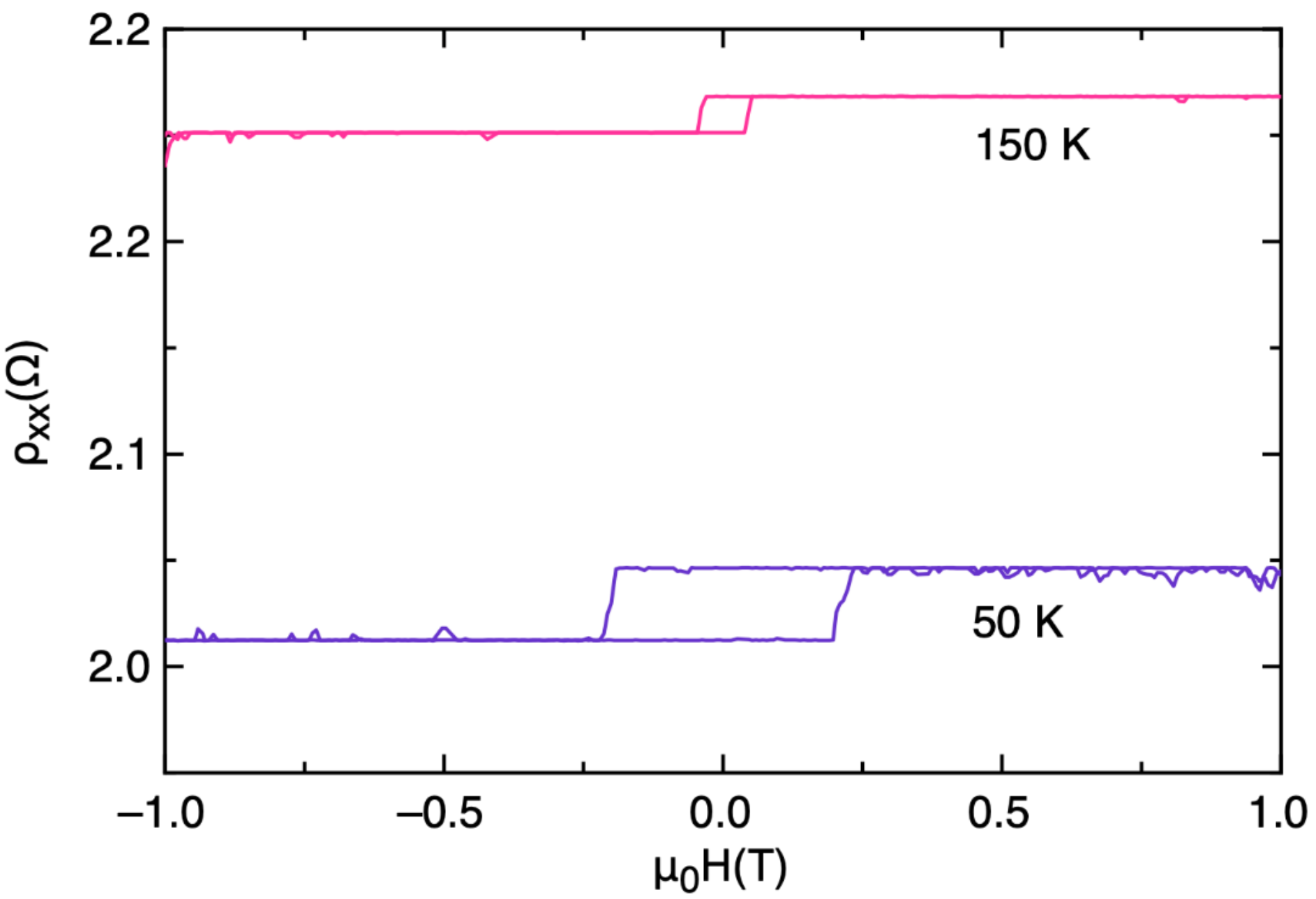

Fig. S4: $\rho_{xx}$ shows similar square loop as $\rho_{xy}$ in 60L FGT.

**E. Field dependent magnetic domain for FGT and $WTe_2$/FGT samples**

Here we show how the magnetic domains in $WTe_2$/FGT differ from FGT in the Fig. S5. For FGT without $WTe_2$, the magnetization saturates and it enters ferromagnetic phase when the field is $660\ Oe$. While a group of skyrmions shows up in the region with $WTe_2$ at $195\ K$. For FGT with $WTe_2$, the DMI penetrates a depth at the interface and disappears for FGT away from the interface. But for the FGT away from the interface, it enters uniform ferromagnetic phase and contributes no contrast. Thus the image captured for $WTe_2$/FGT is with the skyrmions at the interface.

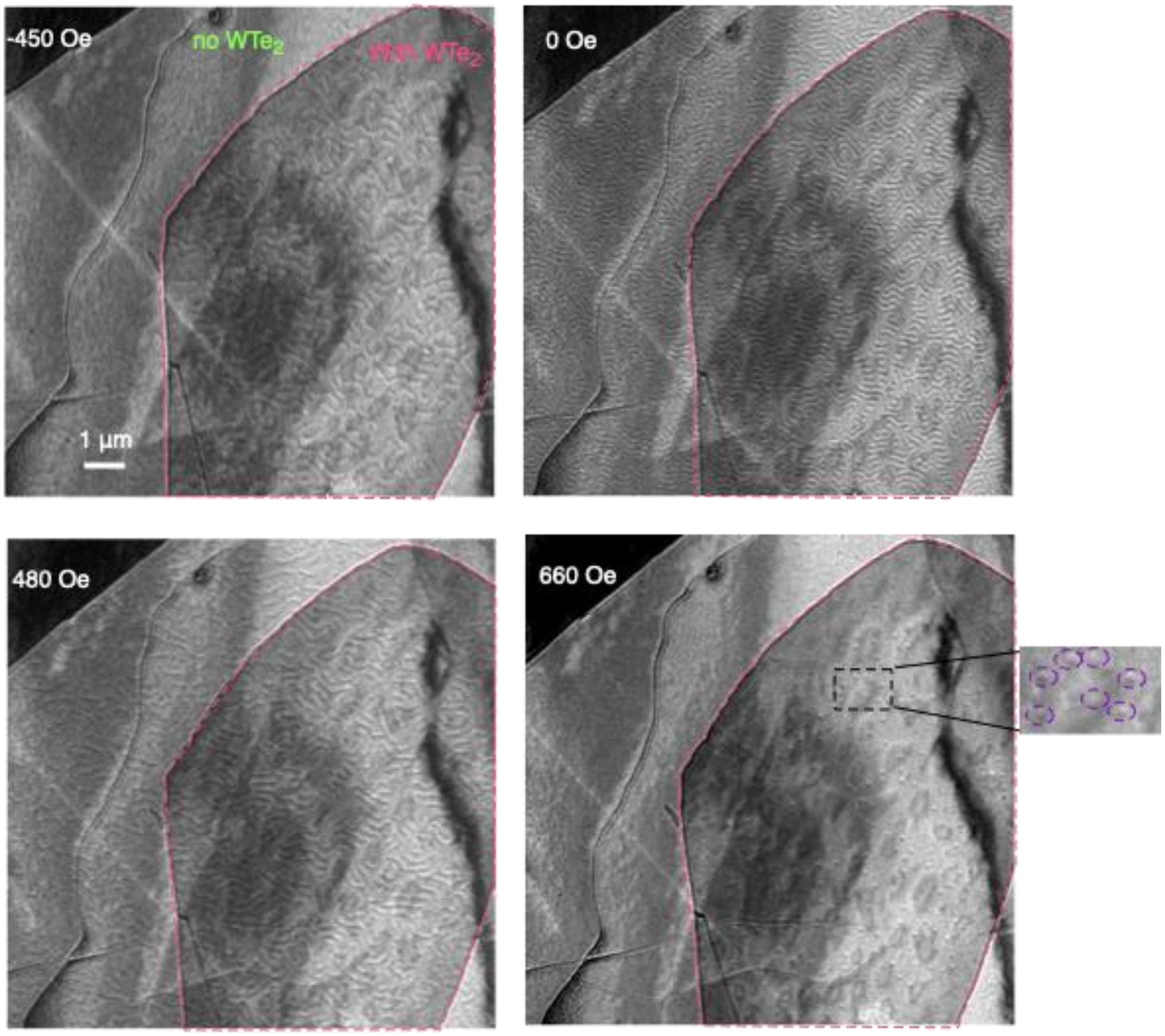

Fig. S5: Magnetic domain difference for 35L FGT with and without 8L $WTe_2$ at $195\ K$ with a tilting angle $\alpha = 30°$ and a varied field. The dashed pink line region is FGT with $WTe_2$. We've zoomed in and indicated the skyrmions with purple dashed circles.

## F. DMI at the interface of $WTe_2$ and FGT

We assume the DMI mainly occurs at the interface between $WTe_2$ and FGT. This is supported by our L-TEM data shown in Fig. S6. When the FGT layer is 35 L, the stripe domain period is smaller compared to the regions without the $WTe_2$ capping, as shown in Fig. S6a. However, when the FGT layer is ~65 L, no observable difference was captured, as shown in Fig. S6b. Besides, when the thickness of FGT is reduced to 30L, the interface plays a more important role, resulting in denser stripe domains in the $WTe_2$ capped regions, as shown in Fig. 4 in the main text. This indicates the DMI is more pronounced in thinner FGT systems with $WTe_2$. Thus the DMI from the interface can penetrate a finite depth into FGT.

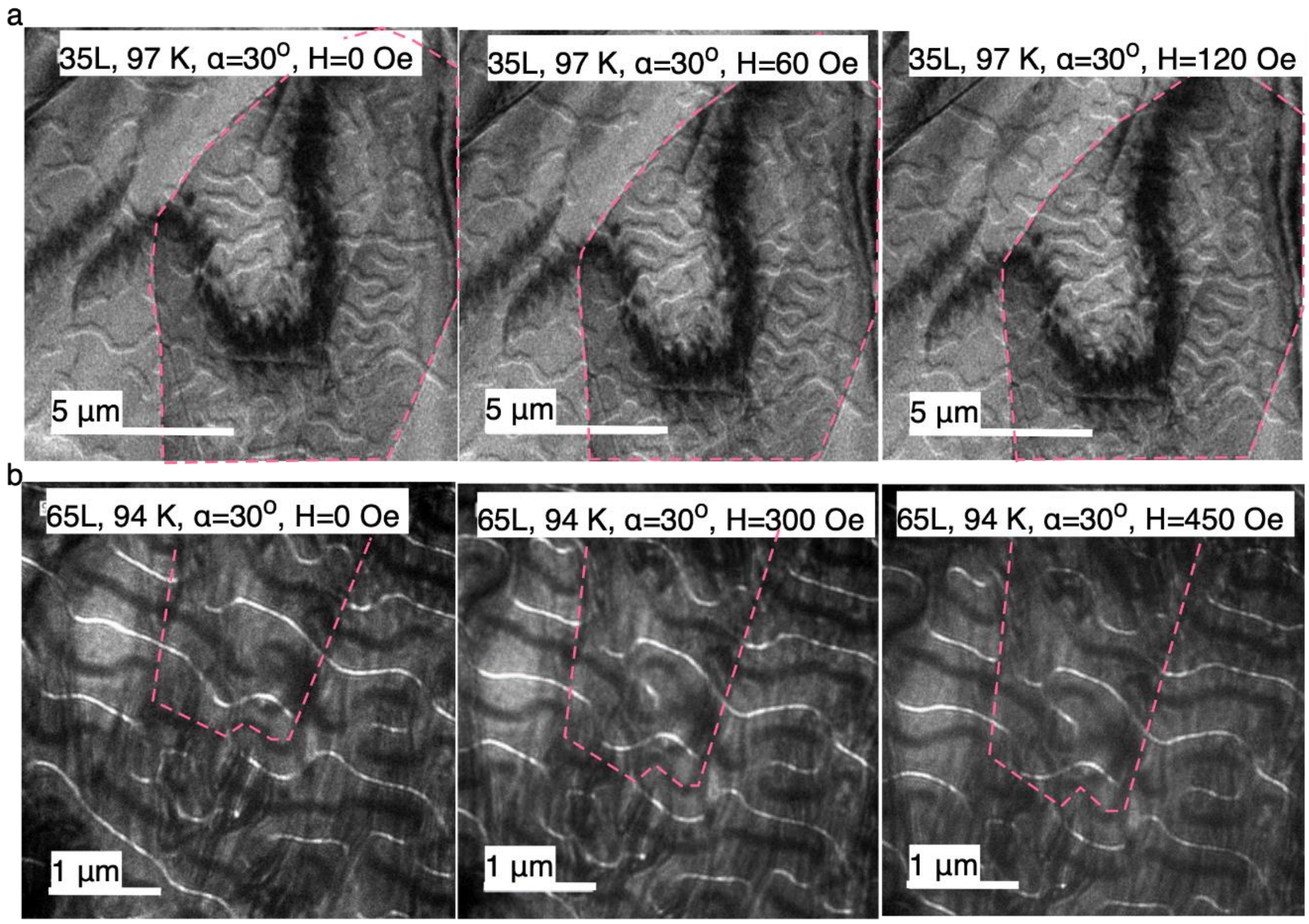

Fig. S6: Magnetic domain difference between FGT and $WTe_2$/FGT samples. a, For 35L FGT, the region with $WTe_2$ shows narrower domain width. b, For 65L FGT, there's no magnetic domain difference. The region with dashed line is for FGT with $WTe_2$.

## G. Focus change during the L-TEM measurements

we checked the skyrmion lattice from under focus to over focus for $WTe_2$/40L FGT at 180 $K$. Skyrmions are only observed at de-focused images. As shown in Fig. S7, under and over focused L-TEM images exhibit the opposite dark-bright color contrast, suggesting Néel-type skyrmions.

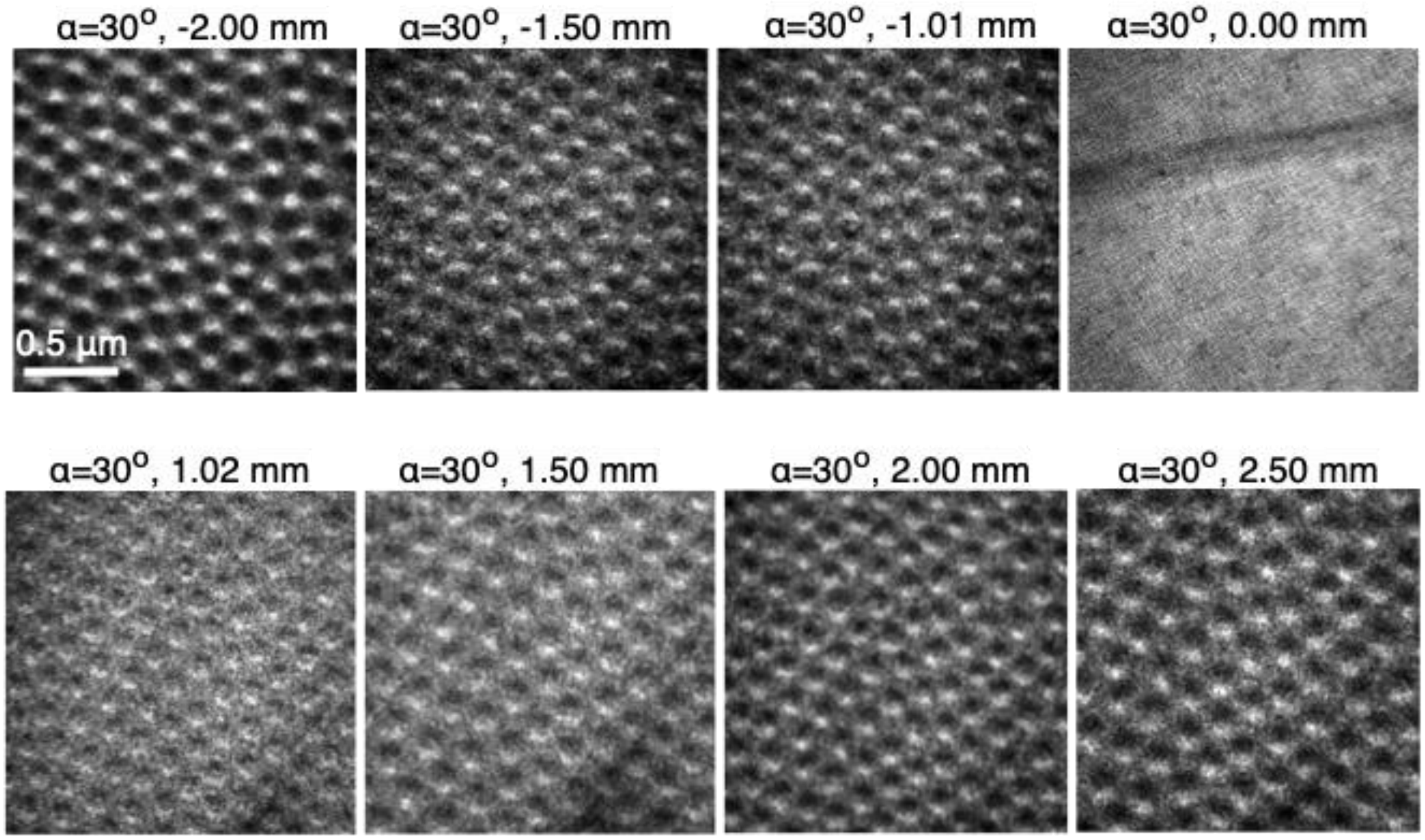

Fig. S7: Focus was changed from under focus to over focus on $WTe_2$/40L FGT samples with a field of 510 $Oe$ at 180 $K$.

## H. Skyrmion size

As shown in Figure S8, a line profile is used to analyze the contrast for a skyrmion. The distance between the lowest and highest data points is the skyrmions size.

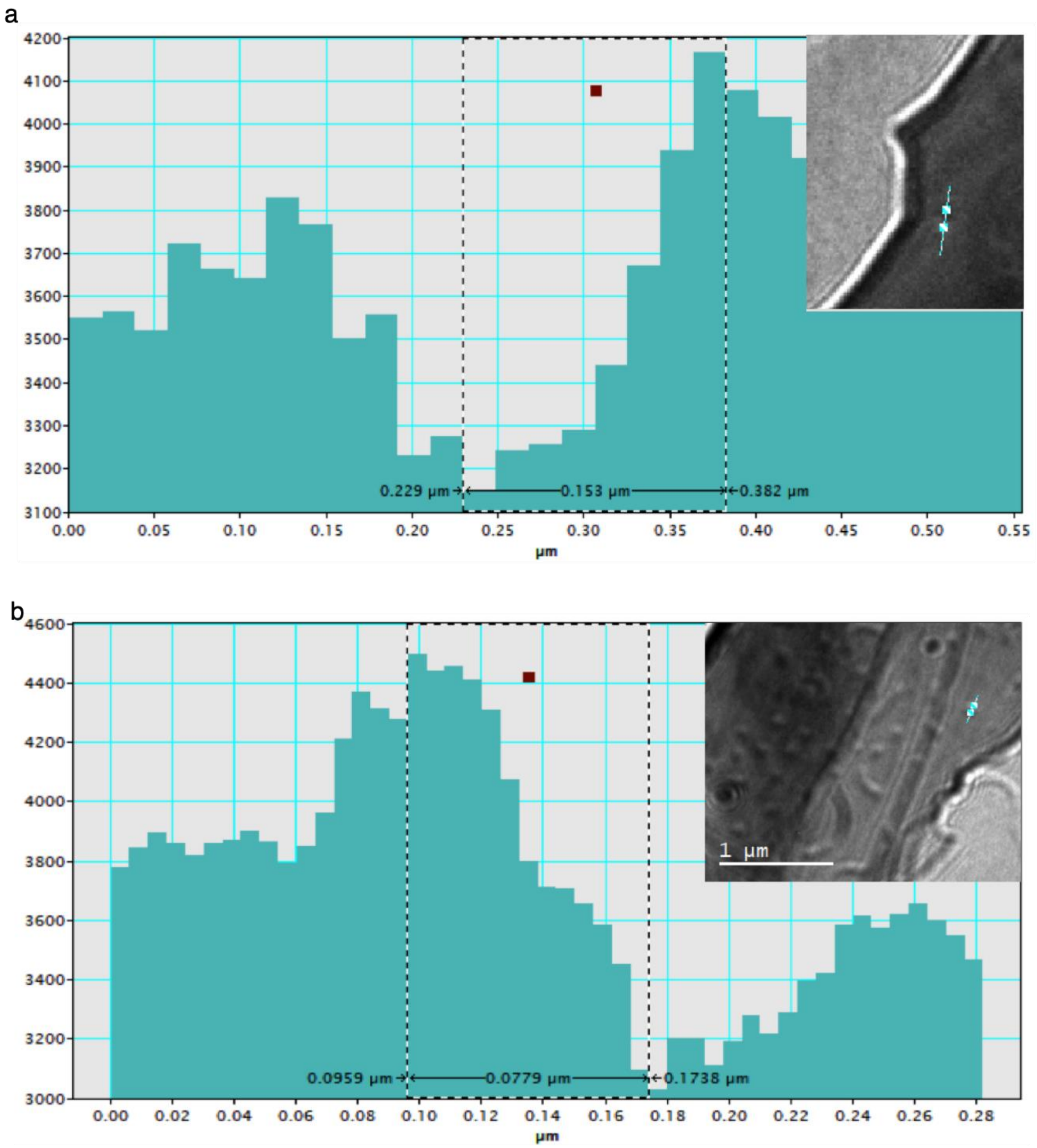

Fig. S8: Line profile for the image of skyrmions observed in 2L $WTe_2$/30L FGT samples. a, The skyrmion size is determined to be $\sim 150\ nm$ at $94\ K$. b, The skrymion size is determined to be $\sim 80\ nm$ at $198\ K$.

## I. Measurement of domain width

Bodenberger and Hubert[3] used a stereological method to define the surface magnetic domain width $w$ of complicated or arbitrary magnetic structure patterns. In their method, $w$ is defined as:

$$w = \frac{2 \times total\ test\ line\ width}{\pi \times number\ of\ intersections}, \quad (1)$$

which appears to be the most universal and commonly applied method[4-6]. In this method, an effective domain width is defined as the ratio of total test line length to the number of intersections of domain walls. For the purpose of evaluating the total domain width, four test straight lines running in random directions is used; the method is illustrated in the image of Fig. S9, where four test lines are drawn. The determined domain width is $290 \pm 10\ nm$.

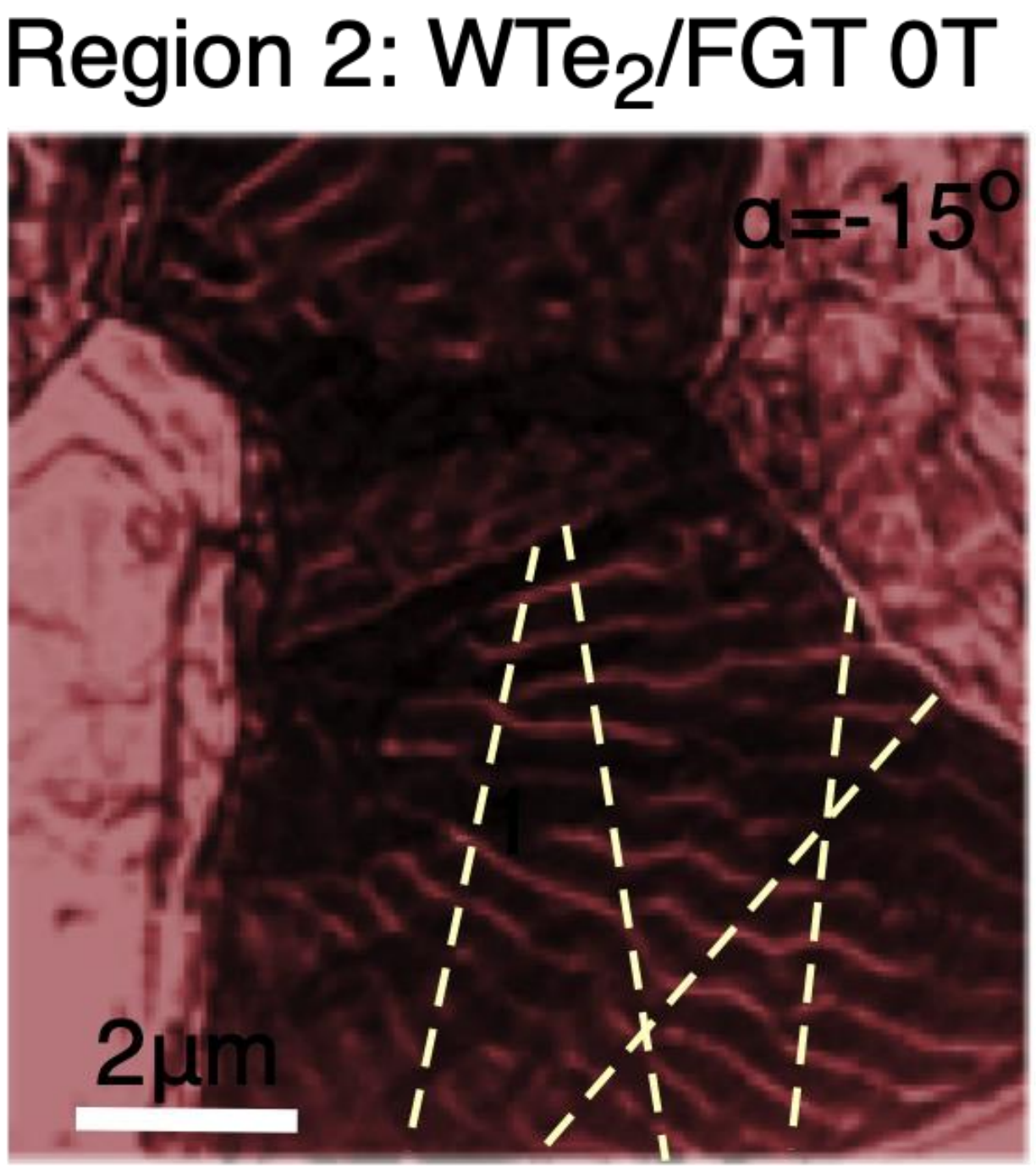

Fig. S9: Representative image used to obtain the average domain size of $WTe_2$/30L FGT sample.

**J. Estimation of DMI constant**

Based on the Stoner-Wohlfarth model[7],the uniaxial anisotropy constant $K_u$ can be

derived via:

$$\frac{2K_u}{M_s} = \mu_0 H_{sat}. \quad (2)$$

As shown in Fig. S10 for 2L $WTe_2$/30L FGT heterostructure, $H_{sat}$ decreases when the temperature increases. Thus we can determine the ratio of the uniaxial anisotropy constant at $5\ K$ $K_{u-5K}$ and at $94\ K$ $K_{u-94K}$. Meanwhile, Ref[8] lists the parameters for bulk FGT around $5\ K: M_{s-5K} = 376\ emu/cm^3$, $K_{u-5K} = 1.46 \times 10^7\ erg/cm^3$, $A = 10^{-7}\ erg/cm$. Thus. $K_{u-94K} = 9.7 \times 10^6\ erg/cm^3$. Since $K_d \ll K_u$, the effective anisotropy constant $K_{eff} \sim 9.7 \times 10^6\ erg/cm^3$. As a result, the domain wall energy for FGT without the DMI contribution is $\sim 3.9\ mJ/m^2$. A DMI constant of $1.0\ mJ/m^2$ is obtained in our system from Equation 2 in the main text.

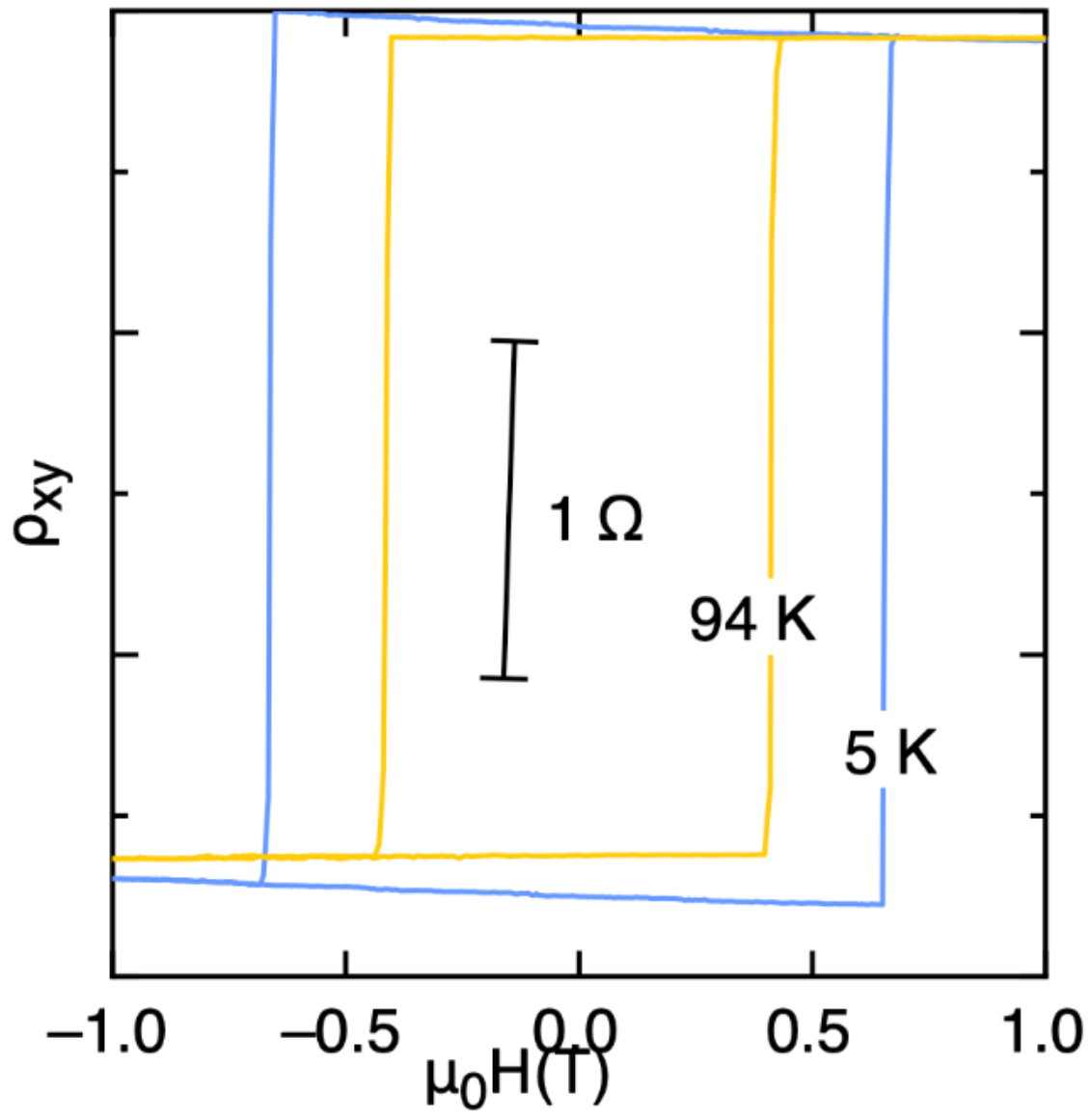

Fig. S10: Hall resistivity of 2L $WTe_2$/30L FGT heterostructure at $5\ K$ and $94\ K$.

**K. Consistency between transport and L-TEM results**

At the humps of $\rho_{xy}$, the total Hall resistivity contains three parts: $\rho_{xy} = \rho_{xy}^{N} + \rho_{xy}^{AHE} + \rho_{xy}^{T}$, where $\rho_{xy}^{N}$ is the normal Hall resistivity, $\rho_{xy}^{AHE}$ is the anomalous Hall resistivity and $\rho_{xy}^{T}$ is the topological Hall resistivity. Assuming a square loop of anomalous Hall effect ($\rho_{xy}^{AHE} = \rho_{xy}^{saturated}$) and a linear $\rho_{xy}^{N}$ at the background, we have: : $\rho_{xy} - \rho_{xy}^{saturated} = \rho_{xy}^{N} + \rho_{xy}^{T} = \frac{1}{ne}(B + B_{eff})$. Here, the topological Hall effect is attributed to an effective field, $B_{eff}$. Since each magnetic skyrmion contributes a flux quantum, $\Phi_0$, assuming a uniform hexagonal skyrmion lattice, we have $B_{eff} = \frac{\Phi_0}{\frac{\sqrt{3}}{2}r^2}$, where $r$ represents the skyrmion lattice constant or skyrmion size. The Hall coefficient $\frac{1}{ne}$ can be further obtained from the slope of $\rho_{xy}$ after magnetic saturation. The skyrmion size can therefore be estimated as $r = \sqrt{\frac{\Phi_0}{\frac{\sqrt{3}}{2}[(\rho_{xy} - \rho_{xy}^{saturated})ne - B]}}$.

Several more $WTe_2$/FGT heterostructures with varied FGT thicknesses show topological Hall loops in Fig. S11a. From the topological Hall effect, we obtained the skyrmion lattice constant from these transport signatures at $100\ K$, as shown by the red circles in Fig.S11b. It is recognized that such estimation oversimplifies the spin texture and can only provide order-of-magnitude estimation. Fortunately, we have obtained a well-resolved skyrmion lattice in thick (40L) FGT samples at $180\ K$. The L-TEM observed skyrmion sizes are illustrated by the squares in Fig. S11b, where the colors of the squares denote the temperature. Unfortunately, observing domain structures by L-TEM in thin FGT samples still fails in our experiment. However, we do see the skyrmion sizes obtained by the two methods fall into the same order of magnitude.

On the other side, as shown in Fig. S6, the $WTe_2$ capping can only impact the domain structure for < 65L FGT films, suggesting a vertical profile of the DMI. Assuming an exponential decay in the DMI profile, our simulation suggests that the skyrmions can only penetrate to a finite depth, where a large volume of ferromagnetic phase shows up away from the interface, as shown in Fig. S12. As discussed before, due to frequent scattering, when carriers are passing through the ferromagnetic phase, they quickly lose the memory of the transverse velocity provided by the topological Hall effect, and therefore the anomalous Hall effect dominates. This explains the missing topological Hall effect in $\rho_{xy}$ humps in thicker films.

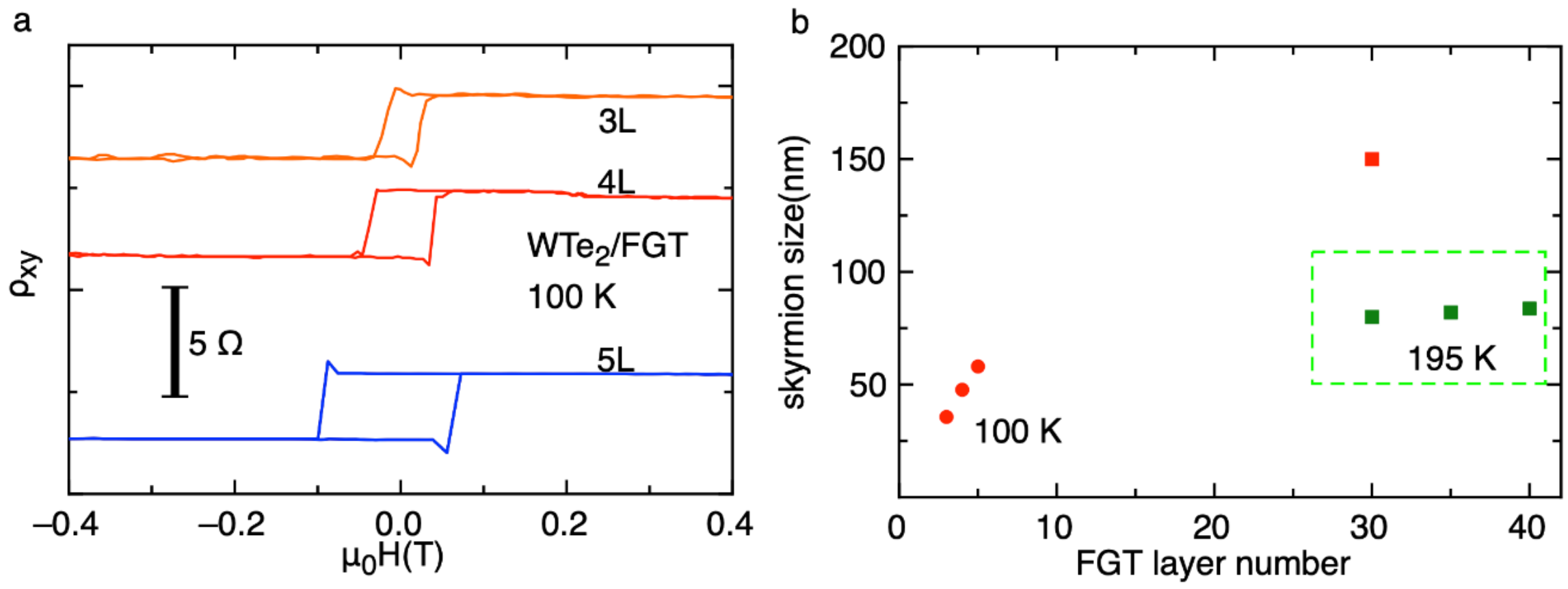

Fig. S11: FGT thickness dependence of skyrmion size in $WTe_2$/FGT heterostructure. a, Topological Hall effect for 1L $WTe_2$/3L FGT, 1L $WTe_2$/4L FGT and 2L $WTe_2$/5L FGT. b, Extracted skrymion size from transport and L-TEM data as a dependence on FGT thickness at $100\ K$ and $195\ K$. The points in a circular shape is the skyrmion size from topological Hall effect and in a square shape is those from L-TEM. Points in red are taken at $100\ K$ and points in green are taken at $195\ K$.

## L. The micromagnetic simulation

The simulation is carried out on a 3D lattice model, with the Hamiltonian written as

$$H = \sum_{<i,j>}[-J\boldsymbol{S_i}\cdot\boldsymbol{S_j} + \boldsymbol{D}_{i,j}(z)\cdot(S_i\cdot S_j)] - \mu_0\sum_i S_i\cdot H_{app}, \quad (3)$$

where $J$ is the Heisenberg exchange coupling, $H_{app}$ denotes the applied magnetic field, and the position-resolved DMI $\boldsymbol{D}_{i,j}(z)$ is given by

$$\boldsymbol{D}_{i,j}(z) = (\hat{\boldsymbol{z}}\times\hat{\boldsymbol{r}}_{i,j})D(z)$$

where $D(z) = D_0 \exp\left(\frac{z-t}{l_0}\right)$. Here, $t$ is the thickness of the film and $l_0$ is a phenomenological penetration depth. The simulation results shown in Fig. 3 in the main text and Fig. S12 are carried out on a cubic lattice defined on a $25 \times 25 \times 20$ mesh. The side walls of the mesh are assumed to be periodic boundaries for simplicity. To mimic the case of a thin film, the top and the bottom surfaces are open, that is, to enforce $\boldsymbol{S}(r) = 0$ for both $z > t$ and $z < 0$. The dynamical behavior of the local spins $\{\boldsymbol{S}_i\}$ follow the Landau Lifshitz Gilbert equation

$$\dot{\boldsymbol{S}} = -\gamma \boldsymbol{S} \times \boldsymbol{H}_{eff} + \alpha \boldsymbol{S} \times \dot{\boldsymbol{S}}, \tag{5}$$

where $\gamma = \frac{g}{\hbar}$ is the gyromagnetic ratio and $\alpha$ is the damping factor. The effective field $\boldsymbol{H}_{eff}$ is given by $\boldsymbol{H}_{eff} = -\frac{\partial H}{\partial \boldsymbol{S}} + \boldsymbol{L}$, where $H$ is the Hamiltonian given by Eq. 3, and $\boldsymbol{L}$ is a random field provided by the thermal fluctuation at finite temperature. The dissipation-fluctuation relation $\langle L_\mu(\boldsymbol{r}, t) L_v(\boldsymbol{r}', t') \rangle = \xi \delta_{\mu,v} \delta_{\boldsymbol{rr}'} \delta_{tt'}$ is satisfied, where $\xi = \frac{\alpha k_B T}{\gamma}$, which is determined by the damping factor and the temperature, $T$, and the average $\langle \dots \rangle$ is taken over the realizations of the fluctuation field. During the simulation, the applied magnetic field sweeps as a triangle wave, with the slopes much smaller than the characteristic time of the spin dynamics, mimicking an adiabatic scan of the applied field in the experiment. The parameters used in this simulation are $\frac{D_0}{J} = 1$, $k_B T = 0.1J$, and $l_0 = t ln(\frac{D_0}{D_{btm}})$, where $D_{btm} = D(z)|_{z=0}$, which is phenomenologically chosen as $D_{btm} = 0.4 D_0$.

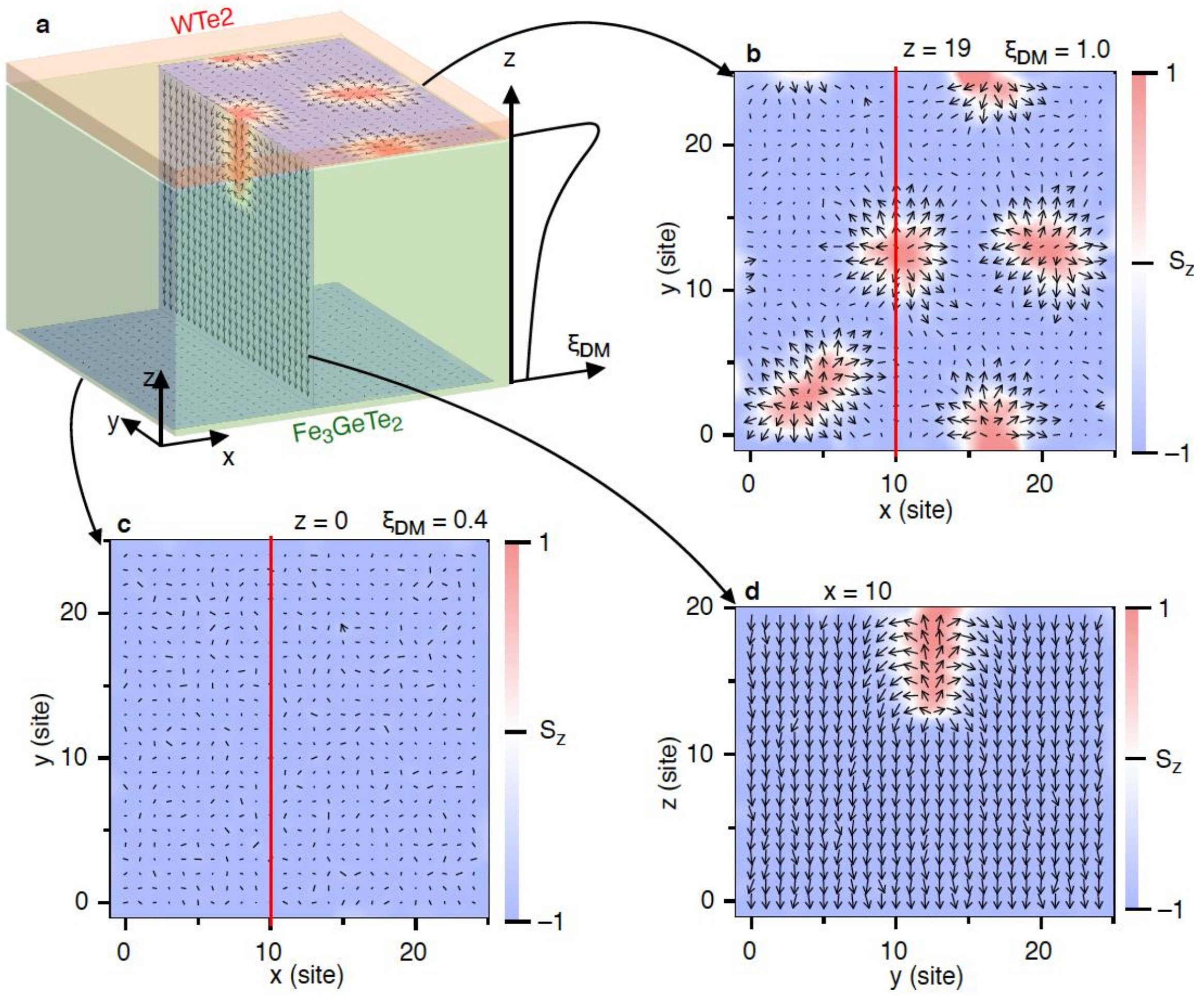

Fig. S12: 3D view from the simulation of the skyrmions in $WTe_2$/FGT. a, DMI exists at the interface between $WTe_2$ and FGT and decays when away from the interface. b, Spin polarization at the interface of $WTe_2$ and FGT. c, Spin polarizations for the side of FGT close to SiN substrate. d, Spin polarization at $yz$ plane with a fixed $x = 10$.